\documentstyle[11pt,fullpage,psfig]{article}
\let\referencesold=\thebibliography
\let\endreferencesold=\endthebibliography
\catcode`@=11
\input prabib.sty
\catcode`@=12
\let\thebibliography=\referencesold
\let\endthebibliography=\endreferencesold
\newif\ifpreprintsty
\begin{document}

\begin{flushright}
UFIFT-HEP-98-16\\
  hep-th/9807151\\
\end{flushright}
\vskip 3.0cm

\renewcommand{\thefootnote}{\fnsymbol{footnote}}
\def\footnoterule{\kern-3pt \hrule width \hsize \kern2.5pt}
\def\ie{{\it i.e.\ }}
\def\eg{{\it e.g.\ }}
\def\beq{\begin{equation}}
\def\eeq{\end{equation}}
\def\beqn{\begin{eqnarray}}
\def\eeqn{\end{eqnarray}}
\pagestyle{empty}

\begin{center}
{\Large\bf Quantum Newtonian Dynamics on a Light Front\footnote
{Supported in part by
the Department of Energy under grant DE-FG02-97ER-41029}
}

\vskip 2.5cm
\large{Charles B. Thorn}\footnote{E-mail  address: thorn@phys.ufl.edu}
\vskip 0.5cm
{\it Institute for Fundamental Theory\\
Department of Physics, University of Florida\\
Gainesville, FL 32611-8440}
\end{center}
\vspace{1.2cm}
\begin{center}
{\bf ABSTRACT}
\end{center}
\noindent We recall the special features of quantum dynamics 
on a light-front
(in an infinite momentum frame) in string and field theory. 
The reason this approach is more effective for string than
for fields is stressed: the light-front dynamics for string
is that of a true Newtonian many particle system,
since a string bit has a fixed Newtonian mass. 
In contrast, each particle of a field theory
has a variable Newtonian mass $P^+$, so the Newtonian 
analogy actually requires an infinite number of species
of elementary Newtonian particles. 
This complication substantially weakens the value of 
the Newtonian analogy in applying light-front dynamics 
to nonperturbative problems. Motivated by the fact that 
conventional field theories can be obtained as infinite tension
limits of string theories, we propose a way to recast field
theory as a standard Newtonian system. We devise and analyze 
some simple quantum mechanical systems that display the essence 
of the proposal, and we discuss prospects for applying these ideas 
to large $N_c$ QCD.
\vskip1in
\begin{flushright}
{Copyright 1999 by The American Physical Society}
\end{flushright}
\vfill
\newpage
\def\balpha{\mbox{\boldmath$\alpha$}}
\def\bgamma{\mbox{\boldmath$\gamma$}}
\def\bsigma{\mbox{\boldmath$\sigma$}}
\def\bepsilon{\hbox{\twelvembf\char\number 15}}
\def\Nlarge{N_c\rightarrow\infty}
\def\Tr{{\rm Tr}}
\newcommand{\ket}[1]{|#1\rangle}
\newcommand{\bra}[1]{\langle#1|}
\newcommand{\firstket}[1]{|#1)}
\newcommand{\firstbra}[1]{(#1|}
\setcounter{footnote}{0}
\pagestyle{plain}
\pagenumbering{arabic}
\renewcommand{\theequation}{\thesection.\arabic{equation}}
\section{Introduction}
The possibility of developing the quantum dynamics of
a relativistic system in light-front form has been under
occasionally active investigation since Dirac first 
suggested the idea 50 years ago \cite{diracfront}. In the sixties
simplifications in the derivation of current algebra sum
rules occurred \cite{fubinifsum,dashengmsum} in
an infinite momentum frame, the light-front in another
guise. Studying Feynman diagrams in such a frame,
Weinberg \cite{weinbergfront} discovered much simplified 
rules in which the energies in the
denominators of old-fashioned perturbation theory took the
non-relativistic form $({\bf p}^2+\mu^2)/2P_L$ 
where ${\bf p}$ is the
momentum transverse to the longitudinal momentum $P_L$ taken
to infinity. Later Susskind \cite{susskindgal} systematized
these simplifications by identifying a Galilei invariance 
on this transverse space in which the longitudinal momentum
played the role of Newtonian mass. Thus the essentially
Newtonian character of light-front dynamics was recognized.

Another study of light-front quantum dynamics \cite{bjorkenks} was
inspired by the physics of deep inelastic lepton
scattering, which probes the light-cone singularities of current
correlators. This line of thought leads to an intuitively appealing
description of the parton wave functions \cite{lepageb}. 
For nonperturbative purposes, the utility of these ideas in field theory, 
beyond conceptual clarifications, was limited by the fact that
the ``Newtonian mass'' $P_L$ was actually a continuous
variable ranging from 0 to $\infty$. Furthermore, the standard
vertices of Feynman diagrams gave nonzero amplitudes
for a ``particle'' of Newtonian mass $P_L$ to transform into several
``particles'' with masses $P^k_L$ as long as
$\sum_k P_L^k=P_L$. Thus the Newtonian analogy was imperfect:
a continuously infinite number of species of Newtonian
particles, which could transmute into each other, was required.
A further annoyance is that field modes with $P_L=0$ have no
Newtonian interpretation at all and must be explicitly
removed, either by deleting them or by ``integrating them out,''
before the Newtonian analogy can be exploited. Neither procedure
is without controversy.

It was not until these ideas were applied \cite{goddardgrt} 
to the Nambu-Goto relativistic string \cite{nambugoto}, 
that the full power of the Newtonian
analogy was realized. With light-front time $\tau=x^+=(x^0+x^1)/\sqrt2$
taken as the analogue of Newtonian time and with the points on the string
parameterized so that the density of longitudinal momentum $P^+
\equiv (P^0+P^1)/\sqrt2$ is constant,
the dynamics of relativistic string is identical to that of
ordinary elastic non-relativistic string moving and vibrating
in the transverse space, described by coordinates $x^k$, $k=2,\ldots D-1$. 
In this description all information about the motion of string
in the remaining direction $x^-$ is redundant save for its
conjugate momentum $P^+$ which measures the total Newtonian mass
of the string. From this point of view, 
the continuous variability of the Newtonian
mass simply reflects the property that string is made up of continuous
material. It is natural to suppose that in reality, just as with
a violin string, relativistic string is not continuous but made up 
of tiny constituents \cite{gilest},  
string-bits \cite{thornmosc}. With this proposal, the dynamics
of fully interacting string can be formulated as those of a standard
Newtonian system. 

As we have noted above, the light-front description of an ordinary
quantum field theory requires the introduction of Newtonian
``particles'' with every possible value of the mass. This is not
necessary with string because variation in Newtonian mass is naturally
achieved by the breaking and joining of pieces of string containing
various numbers of string-bits. Long ago in pursuit of a
connection between field theory and string theory, we showed that
light-front field theory can be made more ``Newtonian'' by
discretizing the $P^+\rightarrow Mm$ each field quantum can 
carry \cite{thornfishnet}, see also \cite{discretelc}\footnote{
Since these early proposals, a major industry, known 
as Discrete Light Cone Quantization (DLCQ), has developed
from them, starting with
\cite{brodskyp}. The literature in this field is now enormous 
and can be tapped by consulting the recent review article
\cite{brodskyppreport}.}. Thus instead 
of a continuous infinity of species of particles, 
there is only a discrete infinity, one species for each
number $M$ of fundamental mass units $m$. Field theoretic
interactions would then occur in two fundamentally different
ways: (1) There could be Newtonian-like potentials, either
``contact'' delta function potentials, due to quartic
local terms in the original Hamiltonian, or non-local 
potentials induced by integrating out $P^+=0$
modes and/or constrained gauge fields, and (2)
transition interactions in which mass is redistributed either
through exchange in a 2 to 2 vertex or through fission
in 1 to 2 and  1 to 3 vertices or through fusion in 2 to 1
or 3 to 1 vertices.
Indeed, the light-front Hamiltonian $P^-$ of the 
field theory is precisely that of a second-quantized
many-body system, which includes terms that don't conserve
particle number even though Newtonian mass is conserved. 
The difficulties of dealing with such
a Hamiltonian are comparable to those of dealing with
the standard time-like (in Dirac's language ``instant'') form of the
Hamiltonian, which is why the Newtonian analogy has been less useful
in this situation. 

An important inspiration for this work is the 
new optimism about the tractability of `t Hooft's
large $N_c$ limit of $QCD$ \cite{thooftlargen} generated by the
intriguing conjecture that large $N_c$ gauge 
theories are equivalent to classical string theories on certain
Anti-de Sitter backgrounds \cite{maldacena,gubserkp,wittenholog}. 
Even as these ideas are being vigorously
pursued, we think it is important to reconsider
earlier efforts to connect large $N_c$ gauge theory
to string theory. This is especially true since
the status of the conjecture at finite `t Hooft coupling
$N_cg^2$ is problematic, so alternative
ideas might yield useful insight on this score. 
Some twenty years ago, following
a suggestion by `t Hooft \cite{thooftlargen}, we sought
to identify the sum of  planar diagrams, parameterized on
a light-front, with the path integral over a light-front
parameterized world sheet \cite{thornfishnet,gilesmt}\footnote{
For a gauge theory like $QCD$ the validity of such an identification
\cite{browergt} was clouded by the uncertainty of how to effectively
deal with the $P^+=0$ singularities of light-front gauge. We hope
that the ideas advanced in this article will lead to a clarification
of such issues.}.
We found that such an identification 
made sense only in a certain large `t Hooft coupling limit, $N_cg^2\to\infty$,
which enforced a ``wee parton'' approximation.
Interestingly, this is also the limit in which the
Maldacena conjecture has the strongest support, because then
the problematic string theory on a curved AdS background
can be replaced by its well understood low energy
supergravity limit. Away from this
limit it was also clear from our earlier work 
that the light-front approach to 
large $N_c$ field theory 
dictated several physical modifications of the minimal
Nambu-Goto dynamics,
including summing over ``holes'' or ``tears'' in the world sheet and
also over the contribution of ``valence'' partons carrying
a finite fraction of the string momentum. The first complication 
can be neatly handled by simply replacing a harmonic nearest neighbor
wee parton interaction with a short range attractive potential
\cite{thornfock,thornweeparton}. However,
we offered no such efficient way of including valence partons except by
brute force summation. We are therefore motivated to ask whether valence
partons can be effectively included
in the context of a conventional Newtonian many body system
made up of wee partons only.

Thus, the purpose of this article is to explore the possibility
that underlying the light-front form of quantum field theory is a completely
standard Newtonian system of ``bits'' living on the transverse space. 
The fact that perturbative string
dynamics {\it is} Newtonian in this pure sense and, 
in the infinite tension limit, can be described by an
effective quantum field theory, implies that such an
underlying system should, at least in theory, exist:
first obtain string theory from a string-bit model and
then take its infinite tension limit.
Whether it is possible to spell out its dynamics in
a useful way, and whether its existence is any help in
dealing with interesting non-perturbative issues, such as quark
confinement in QCD, are issues we will not address. 
Our aim here is the more modest one of
examining the features such an underlying theory must
possess and using some simple quantum mechanical models to illustrate
how the mechanisms can work.

Our basic proposal is that just as string can be regarded as a
polymeric bound state of string bits, a field quantum can be
regarded as a very tightly bound state of bits. The quantity
of $P^+$ such a quantum carries is just proportional to the
number of bits it contains. If such an interpretation is
successful, string theory and quantum field theory would
be effective low energy descriptions of a single kind of
underlying theory. From a pragmatic standpoint, rephrasing
complicated dynamical issues in quantum field theory, 
such as quark confinement, into a question about
the properties of various kinds of Newtonian many-body
systems could lead to new insights as well as to new
quantitative results.

In the next section we recall how field theoretic interactions
look on the light-front by examining a cubic scalar field interaction.
We then go on in Section 3 to study how the ideas
sketched above play out for a simple 2-bit truncation of
the scalar field model. We exhibit and solve a simple
two particle potential model which serves as the
underlying Newtonian model for the truncated field theory.
The final section is devoted to a discussion of
the prospects for applying these ideas to full-fledged field theory
models, especially large $N_c$ QCD.

\section{The Cubic Vertex in Scalar Field Theory}
\label{sec2}
%
\setcounter{equation}{0}
Let us begin by reviewing the light-front description of a
scalar field. It can be summarized by writing
\begin{equation}
\phi({\bf x}, x^-)=\int_0^\infty {dP^+\over\sqrt{4\pi P^+}}
(a({\bf x},P^+)e^{-iP^+x^-}+a^\dagger({\bf x},P^+)e^{iP^+x^-})
\end{equation}
where $[a({\bf x},P^+),a^\dagger({\bf y},Q^+)]
=\delta({\bf x}-{\bf y})\delta(P^+-Q^+)$. The free field 
Hamiltonian is just
\begin{equation}
H_0=P^-_0=\int_0^\infty d{\bf x}dP^+a^\dagger({\bf x},P^+){(-\nabla^2+\mu^2)
\over2P^+}a({\bf x},P^+).
\end{equation}
A typical field theoretic interaction, a $g\phi^3/6$ term, has
the light front presentation:
\begin{eqnarray}
V_3&=&{g\over8\sqrt\pi}
\int d{\bf x}\int_0^\infty{dP^+dQ^+\over\sqrt{P^+Q^+(P^++Q^+)}}
\nonumber\\
& &\mbox{}\qquad [a^\dagger({\bf x},P^++Q^+)a({\bf x},P^+)a({\bf x},Q^+)+
a^\dagger({\bf x},P^+)a^\dagger({\bf x},Q^+)a({\bf x},P^++Q^+)].
\end{eqnarray}
We would like to explore the possibility that $a^\dagger(P^+)$ creates
not an elementary quantum with Newtonian mass $P^+$, but rather
a tightly bound state of infinitely many bits whose total Newtonian mass
is $P^+$. Begin with a discretization of $P^+=Mm$, where $m$ is
the Newtonian mass of an elementary bit. Then $a({\bf x},P^+)$ is replaced
by $a_M({\bf x})/\sqrt{m}$, so that $[a_M({\bf x}),a^\dagger_N({\bf y})]
=\delta_{MN}\delta({\bf x}-{\bf y})$. Then the preceding equations reduce
to
\begin{equation}
\phi({\bf x}, x^-)=\sum_{M=1}^\infty {1\over\sqrt{4\pi M}}
(a_M({\bf x})e^{-iMmx^-}+a^\dagger_M({\bf x})e^{iMmx^-})
\end{equation}
with the free field Hamiltonian 
\begin{equation}
H_0=P^-_0={1\over m}\int d{\bf x}\sum_{M=1}^\infty 
a^\dagger_M({\bf x}){(-\nabla^2+\mu^2)\over2M}a_M({\bf x}).
\label{freehamiltonian}
\end{equation}
and the cubic interaction
\begin{eqnarray}
V_3&=&{g\over8m\sqrt\pi}\int d{\bf x}\sum_{M,N=1}^\infty{1\over\sqrt{MN(M+N)}}
\nonumber\\
& &\mbox{}\qquad [a^\dagger_{M+N}({\bf x})a_M({\bf x})a_N({\bf x})+
a^\dagger_M({\bf x})a^\dagger_N({\bf x})a_{M+N}({\bf x})].
\label{cubicvertex}
\end{eqnarray}
Note that our discretization includes a prescription for regulating
the notorious $P^+=0$ singularities of light cone quantization: the
$M=0$ terms are simply deleted. We therefore implicitly assume  
that any physical phenomena involving $P^+=0$ are adequately described
as a limit from $P^+>0$. 
This might, of course, require that the modes with $P^+=0$
be ``integrated out,'' inducing new interactions amongst
the modes with $P^+\neq 0$. 
In cases where the $P^+=0$ problems can not be dealt with in this
way (see, for example \cite{mccartor}), the Newtonian analogy
would fall short in an important respect, and the more far-reaching aspects
of our proposal of a perfect Newtonian analogy would not apply.

If $a^\dagger_{M}$ creates a bound state, rather than an elementary
quantum, the interaction (\ref{cubicvertex}) is to be regarded
as a term in an effective Hamiltonian, which reproduces
a transition process in the underlying theory in the limit where
the size of the composite state is negligible compared to the
wavelengths characterizing the transition. The factors 
$1/\sqrt{MN(M+N)}$, crucial for Poincar\'e invariance, 
must arise as properties of the bound
system and are not automatic. For example, in the case
of the discretized bosonic string, it was shown in \cite{gilest}
that the square root is, in generic transverse dimensionality $d$,
replaced by the fractional power $d/48$. This leads to the
critical dimensionality $d=24$. Although string theory provides
an existence proof for an appropriate binding mechanism, we are
suggesting that the phenomenon could be more general.

The free Hamiltonian (\ref{freehamiltonian}) includes a term
giving the free particle energy for each value of $M$.
For $g=0$ the $M$ dependence displayed
is required by Lorentz invariance. If each quantum is in
fact a composite, the energy is given by the binding
dynamics and cannot be put in by hand. However the 
coefficient of $-\nabla^2$ is guaranteed by the underlying
Galilei invariance of this dynamics: the term just gives
the center of mass kinetic energy. The $M$ dependence
of the term $\mu^2/2mM$ is {\it not} guaranteed {\it a priori}
and represents a limitation on the binding dynamics. In 
the case of string viewed as a polymer of string-bits,
this dependence arises for large $M$ due to the 
one-dimensionality of the bound system (so the length
of the system is proportional to $M$) and the 
universal $1/{\rm length}$ dependence of phonon energies.
Notice, for example, that an ordinary {\bf elastic} $p$-brane would
have a linear size proportional to ${M}^{1/p}$ and therefore an
incorrect $M$ dependence unless $p=1$. However, when a relativistic
membrane is viewed on a light front, the restoring
energies are of order $(\partial x)^{2p}$, giving a classical
energy estimate of order $(1/{\rm size})^p$ restoring, at least
superficially, the correct $M$ dependence.

\setcounter{equation}{0}
\section{A Two-Bit Model}
In order to illustrate the manner in which
an effective ``elementary'' quantum with $M\neq1$ may be regarded as a bound
state of quanta with $M=1$ only, we turn to an admittedly
highly rarefied truncation of the scalar field theory described in the
previous section. We specify this model by restricting the
scalar field theory to the sector with $M=2$. That is we have only
two classes of Fock states: those with two quanta with $M=1$ and
those with a single quantum with $M=2$. A general state in
this sector therefore has the representation
\begin{equation}
\ket{\psi,\chi}=\int d{\bf x}_1d{\bf x}_2 a_1^\dagger({\bf x}_1)
a_1^\dagger({\bf x}_2)\ket{0}\psi({\bf x}_1,{\bf x}_2)
+\int d{\bf x} a_2^\dagger({\bf x})\ket{0}\chi({\bf x}).
\label{2bitstate}
\end{equation}
With this truncation the cubic vertex reduces to only two terms
\begin{eqnarray}
V^{Trunc}_3={g\over8m\sqrt{2\pi}}\int d{\bf x}
[a^\dagger_{2}({\bf x})a_1({\bf x})a_1({\bf x})+
a^\dagger_1({\bf x})a^\dagger_1({\bf x})a_{2}({\bf x})].
\label{2bitcubicvertex}
\end{eqnarray}

The time independent Schr\"odinger equation for this system is thus
a coupled pair of differential equations:
\begin{eqnarray}
{1\over2m}[-\nabla_1^2-\nabla_2^2 + 2\mu^2-2mE]\psi({\bf x}_1,{\bf x}_2)
+{g\over8m\sqrt{2\pi}}\delta({\bf x}_1-{\bf x}_2)\chi({\bf x}_1)&=&0
\nonumber\\
{1\over4m}[-\nabla^2+\mu_2^2-4mE]\chi({\bf x})+2{g\over8m\sqrt{2\pi}}
\psi({\bf x},{\bf x})&=&0.
\label{2bitschr}
\end{eqnarray}
Notice that we have allowed for the $M=2$ quantum to have a 
``bare'' Lorentzian mass $\mu_2$ different from that of
the $M=1$ quantum $\mu$. This is because the Lorentzian mass
of the $M=2$ quantum is obviously renormalized by the interactions
unlike that of the $M=1$ quantum.

By Galilei invariance we may work in the center of mass system
for which $\psi$ is a function $f({\bf x})$ only of the relative coordinate
${\bf x}\equiv{\bf x}_1-{\bf x}_2$ and $\chi$ is a constant.
Then the second equation can be trivially solved for $\chi$,
which can then be substituted back into the first equation to
give the single particle Schr\"odinger equation
\begin{eqnarray}
[-\nabla^2 + \mu^2-mE]f({\bf x})
-{g^2\over8{\pi}(\mu_2^2-4mE)}\delta({\bf x})f({\bf 0})=0
\label{relativeschr}
\end{eqnarray}
The delta function potential is of course singular in most
transverse dimensionalities, so we need to regulate it.
A convenient regularization is to specify that 
$\delta({\bf x})$ is replaced by a radial delta function 
$\delta(|{\bf x}|-a)/a^{d-1}\Omega_d$
displaced a distance $a$ from the origin on $s$-waves 
and is zero on all other states.
Here $\Omega_d=2\pi^{d/2}/\Gamma(d/2)$ is the volume of a unit 
$(d-1)$-sphere. Then (\ref{relativeschr}) gives non-trivial 
dynamics on $s$-waves where it reduces to the radial
equation ($f_{s-wave}\equiv R(r)$)
\begin{equation}
\left[-{d^2\over dr^2}-{d-1\over r}{d\over dr} + \mu^2-mE\right]R(r)
-{g^2\over8{\pi}\Omega_d a^{d-1}(\mu_2^2-4mE)}\delta(r-a)R(a)=0
\label{radialschr}
\end{equation}

This simple model can be completely solved. In order that the
$M=2$ quantum have the same Lorentzian mass as the $M=1$
quantum, we require that there be a discrete $s$-wave energy
eigenstate with $E=\mu^2/4m$. This condition will determine
the bare mass $\mu_2$. Putting $\kappa\equiv\sqrt{\mu^2-mE}
=\sqrt{3}\mu/2$, the solutions of (\ref{radialschr}) for $r\neq a$
are the Bessel functions $I_\nu(\kappa r), K_\nu(\kappa r)$
with $\nu=(d-2)/2$ for $r<a$, $r>a$ respectively. Continuity
at $r=a$ and the discontinuity in first
derivatives implied by the delta potential leads to the
relation
\begin{equation}
\mu_2^2=\mu^2+{g^2\over8{\pi}\Omega_d a^{d-2}}
K_{\nu}(\kappa a)I_{\nu}(\kappa a)
\end{equation}
Of course $\mu_2$ diverges as $a\to 0$, for $d\geq2$. 

Now we turn to the continuous part of the spectrum with
$\mu^2-mE\equiv -k^2<0$. Then the solutions are the ordinary
Bessel functions $J_\nu(kr), N_\nu(kr)$. The $s$-wave phase shift is
then determined by the matching conditions at $r=a$ of the two forms
\begin{eqnarray}
R(r)&=& J_\nu(kr)\phantom{\cot\delta J_\nu(kr)-N_\nu(kr)\over 
\cot\delta J_\nu(ka)-N_\nu(ka)}\qquad r<a \nonumber\\
R(r)&=& J_\nu(ka){\cot\delta J_\nu(kr)-N_\nu(kr)\over 
\cot\delta J_\nu(ka)-N_\nu(ka)}\qquad r>a. 
\end{eqnarray}
Solving these conditions gives
\begin{equation}
\cot\delta={N_\nu(ka)\over J_\nu(ka)}+{16\Omega_d a^{d-2}(\mu^2-4mE)
\over g^2 J_\nu(ka)^2}+{2\over\pi}{K_\nu(\kappa a)I_\nu(\kappa a)
\over J_\nu(ka)^2}
\end{equation}
Recalling that $a\neq0$ was a temporary regulator, we now take the
limit $a\to0$, which exists at fixed $g$ for $\nu<1$ corresponding to $d<4$,
\begin{equation}
\cot\delta\to \cot\pi\nu-\csc\pi\nu\left({\kappa\over k}\right)^{2\nu}
-\Gamma(1+\nu)^2{2^{2\nu+6}\Omega_d(k^2+\kappa^2)\over g^2k^{2\nu}}.
\label{fieldeffrange}
\end{equation}
We notice that this is just the $d$-dimensional effective range
approximation, 
\begin{equation}
k^{2\nu} \cot\delta=k^{2\nu}\cot\pi\nu-\kappa^{2\nu}\csc\pi\nu
-\Gamma(1+\nu)^2{2^{2\nu+6}\Omega_d(k^2+\kappa^2)\over g^2}.
\end{equation}
The familiar case of $d=3$ corresponds to $\nu=1/2$, whence the
effective range formula is
$$k\cot\delta=-{1\over a_s}+{1\over2}r_{eff}k^2,$$
with $a_s$ the scattering length and $r_{eff}$ the
effective range. Thus the $r_{eff}$ is {\it negative} 
for this system.
Since $d$ is transverse dimensionality, it would actually be 2
in our 3 dimensional world, corresponding to $\nu=0$. In that case
our general formula reduces to
\begin{equation}
\cot\delta={2\over\pi}\ln{k\over\kappa}-{128\pi(k^2+\kappa^2)
\over g^2}\Gamma(1+\nu)^2.
\end{equation}

For this simple system, the question we pose in this article is
whether the system can be equally well described by a two 
$M=1$ particle system, without the explicit introduction of a new
species of particle with $M=2$. To be more precise, we do not mean to
``integrate out'' the $M=2$ field as in (\ref{relativeschr}), which
gives an effective two particle dynamics. The presence of
$E$ dependence in the potential term is the tipoff
that a degree of freedom has been eliminated, and this 
is what we want to avoid. 
In other words, we seek a two particle
potential {\it independent} of $E$ which reproduces the same physics
as (\ref{relativeschr}). Since the effective range approximation
is a universal low energy behavior for potential scattering, we
expect that there are many potentials that do the trick.
However, since a negative effective range is perhaps unfamiliar, we
think it illuminating to exhibit a particular sample potential which 
yields the desired behavior. 

\begin{figure}
\centering
\centerline{
\psfig{figure=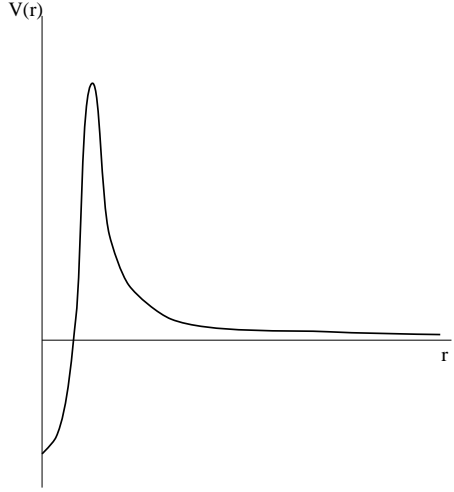,height=6.5cm}
}
\caption{Potential energy function for the two bit model.}
\label{fieldpot}
\end{figure}

To avoid the usual positive sign of the
effective range, it is essential to use a potential that is
not monotonic. A simple tractable choice which does the job is
a potential of the form
\begin{equation}
V(r)=-\gamma\delta(r-b)+\lambda\delta(r-a),\qquad\qquad 0<b<a\quad{\rm and}
\quad \gamma,\lambda>0.
\label{newpotential}
\end{equation}
This is an idealized version of a more generic potential
of the shape shown in Fig. \ref{fieldpot}
The important qualitative features here are an attractive potential
to produce a bound state to simulate the $M=2$ particle, and a repulsive
barrier to suppress the coupling of this bound state to
the two particle state unless the two particles are within 
a distance of $O(a)$ from each other. In the limit $a\to0$,
the couplings can be tuned so that the physics of (\ref{relativeschr})
is reproduced.

Here is a sketch of the calculation. The $s$-wave radial wave function
is given in the three regions by
\begin{eqnarray}
R(r)&=& J_\nu(kr)\phantom{{\cot\phi J_\nu(ka)-N_\nu(ka)\over 
\cot\phi J_\nu(kb)-N_\nu(kb)}\ 
{\cot\delta J_\nu(kr)-N_\nu(kr)\over 
\cot\delta J_\nu(ka)-N_\nu(ka)}}\qquad r<b \nonumber\\
R(r)&=& J_\nu(kb){\cot\phi J_\nu(kr)-N_\nu(kr)\over 
\cot\phi J_\nu(kb)-N_\nu(kb)}\ \phantom{\cot\delta J_\nu(kr)-N_\nu(kr)\over 
\cot\delta J_\nu(ka)-N_\nu(ka)}\qquad b<r<a \nonumber\\ 
R(r)&=& J_\nu(kb){\cot\phi J_\nu(ka)-N_\nu(ka)\over 
\cot\phi J_\nu(kb)-N_\nu(kb)}\
{\cot\delta J_\nu(kr)-N_\nu(kr)\over 
\cot\delta J_\nu(ka)-N_\nu(ka)}\qquad r>a. 
\end{eqnarray}
The jump condition for $R^\prime$ at $r=b$ and $r=a$ can be solved 
for $\cot\phi$ and $\cot\delta$:
\begin{eqnarray}
\cot\phi&=&{N_\nu(kb)\over J_\nu(kb)}+{1\over{\hat \gamma}J_\nu^2(kb)}
\nonumber\\
\cot\delta&=&{N_\nu(ka)\over J_\nu(ka)}+{N_\nu(ka)-J_\nu(ka)\cot\phi\over
 J_\nu(ka)[{\hat\lambda}\{J_\nu^2(ka)\cot\phi-J_\nu(ka)N_\nu(ka)\}-1]},
\end{eqnarray}
where to reduce clutter, we have defined ${\hat\lambda}=\pi m\lambda a/2$
and ${\hat\gamma}=\pi m\gamma b/2$. 
Eliminating $\cot\phi$ in the second of these 
equations and rearranging factors slightly leads to
\begin{equation}
\cot\delta={N_\nu(ka)\over J_\nu(ka)}-{1\over J^2_\nu(ka)}
\left[{\hat\lambda}+{1\over J^2_\nu(ka)}\left(
{N_\nu(ka)\over J_\nu(ka)}-{N_\nu(kb)\over J_\nu(kb)}-
{1\over{\hat\gamma}J^2_\nu(kb)}\right)^{-1}\right]^{-1},
\label{potshift}
\end{equation}

To compare with (\ref{fieldeffrange}), we need to study the low energy
behavior of the phase shift, \ie\ we take $ka<<1$. The small argument
behaviors of the Bessel functions yield
\begin{eqnarray}
{1\over J^2_\nu(z)}&=&\Gamma(1+\nu)^2\left({z\over2}\right)^{-2\nu}\left[1
+{z^2\over2(1+\nu)}+O(z^4)\right]
\nonumber\\
{N_\nu(z)\over J_\nu(z)}&=&\cot\pi\nu-
{\Gamma(1+\nu)^2\over\pi\nu}\left({z\over2}\right)^{-2\nu}
\left[1-{\nu z^2\over2(1-\nu^2)}+O(z^4)\right]
\end{eqnarray}
For definiteness we restrict our low energy analysis to $\nu$
in the range $0<\nu<1$
which will cover the case $d=3$ and the case $d=2$ as a limit. 
Then inspection shows that
the first term of (\ref{potshift}) has a singular behavior as $a\to0$
whose cancelation requires that the quantity within square brackets
approaches $-\pi\nu$. Further, in order to yield a nontrivial phase shift
in the limit, this value must be approached as the power $a^{2\nu}$: 
\begin{equation}
{\hat\lambda}+{1\over J^2_\nu(ka)}\left(
{N_\nu(ka)\over J_\nu(ka)}-{N_\nu(kb)\over J_\nu(kb)}-
{1\over{\hat\gamma}J^2_\nu(kb)}\right)^{-1}=-\pi\nu+O(a^{2\nu}).
\label{limit}
\end{equation}
This can be achieved by tuning the $a$ dependence of ${\hat\gamma},
{\hat\lambda}$. Putting in the small argument expansion for the
Bessel functions in (\ref{limit}) gives, with $\eta\equiv b/a$
\begin{eqnarray}
&&{\hat\lambda}+\left[1+O(a^2)\right]\left(
{\eta^{-2\nu}\over\pi\nu}
\left(1-{\nu k^2b^2\over2(1-\nu^2)}\right)
-{1\over\pi\nu}
\left(1-{\nu k^2a^2\over2(1-\nu^2)}\right)
-{\eta^{-2\nu}\over{\hat\gamma}}\left(1+{k^2b^2\over2(1+\nu)}\right)\right)^{-1}
\nonumber\\
&&=-\pi\nu+O(a^{2\nu}).
\label{limit2}
\end{eqnarray}
In order to have $k$ dependence in the limit, ${\hat\gamma}$ must be
tuned so that quantity in the denominator of the second term vanishes 
as the power $a^{1-\nu}$, so that the quadratic terms in $k$ will
contribute the requisite power $a^{2\nu}$. Thus put
\begin{equation}
{1\over\pi\nu}(\eta^{-2\nu}-1)
-{1\over{\hat\gamma}}\eta^{-2\nu}=-\xi a^{1-\nu}{1-(a/\ell)^{2\nu}\over\nu},
\end{equation}
where the extra factor ensures the proper behavior at $\nu=0$.
Then (\ref{limit2}) becomes
\begin{eqnarray}
&&{\hat\lambda}+\left[1+O(a^2)\right]\left(
-\xi a^{1-\nu}{1-(a/\ell)^{2\nu}\over\nu}
+(ka)^2\left[{1-\eta^{2-2\nu}\over2\pi(1-\nu^2)}
-{\eta^{2-2\nu}\over2{\hat\gamma}(1+\nu)}\right]\right)^{-1}
\nonumber\\
&&\sim{\hat\lambda}-\left[{\nu\over\xi a^{1-\nu}(1-(a/\ell)^{2\nu})}
+O(a^{1+\nu})\right]\left(
1+{k^2a^{1+\nu}\nu\over\xi(1-(a/\ell)^{2\nu})}
\left[{1-\eta^{2-2\nu}\over2\pi(1-\nu^2)}
-{\eta^{2-2\nu}-\eta^{2}\over2\pi\nu(1+\nu)}\right]\right)
\nonumber\\
&&\qquad\qquad=-\pi\nu+O(a^{2\nu}),
\label{limit3}
\end{eqnarray}
where, in the second line, we have substituted the limiting form for
$1/{\hat\gamma}$ in the coefficient of $k^2a^2$, and we have also
approximated the reciprocal by the first two terms of the Taylor series.
We can now easily read off
\begin{equation}
{\hat\lambda}={\nu\over\xi a^{1-\nu}(1-(a/\ell)^{2\nu})}
-{\pi\nu\over(1-(a/\ell)^{2\nu})}
-{C\nu^2a^{2\nu}\over(1-(a/\ell)^{2\nu})^2}
\end{equation}
and thence the $s$-wave phase shift
\begin{equation}
\cot\delta=\cot\pi\nu-{\Gamma(1+\nu)^2\over\pi\nu}
\left({k\ell\over2}\right)^{-2\nu}-{\Gamma(1+\nu)^2\over\pi^2}
\left({k\over2}\right)^{-2\nu}
\left[{C}+{k^2\over\xi^2}\left(
{1-\eta^{2-2\nu}\over2\pi(1-\nu^2)}
-{\eta^{2-2\nu}-\eta^{2}\over2\pi\nu(1+\nu)}\right)\right].
\label{potshift0}
\end{equation}
Notice that the coefficient of $k^2$ will be negative as in
(\ref{fieldeffrange}) if the quantity
$$f(\eta^2)\equiv{\nu(1-\eta^{2-2\nu})\over(1-\nu)}
-{\eta^{2-2\nu}+\eta^{2}}
$$
is positive. To see when this occurs, note that $f^\prime=1-\eta^{-2\nu}<0$
for $0<\eta^2<1$, and $f(0)=\nu/(1-\nu)$, $f(1)=0$. It follows that $f$ is
positive in this interval which is when $b<a$.
We have been careful to set things up so that the case of $d=2$ 
is properly described by the singular limit $\nu\to0$. 

We conclude this section by stating, for this baby two-bit model,
how our results realize the goals set out in the introduction.
The underlying ``microscopic'' theory of the model is the
two particle system described by the potential (\ref{newpotential}).
The parameters of the microscopic theory are the
couplings $\lambda, g$ and the distance scales $a, b$. 
The effective baby field theory is described by the pair
of equations (\ref{2bitschr}), with $g$ the ``bare''
cubic coupling and $\mu,\mu_2$ the ``bare'' Lorentzian
masses. This effective field theory has ultraviolet
divergences which require a regulator. After removing
the regulator, keeping measurable parameters fixed and tuned
so that the ``renormalized'' Lorentzian masses are
the same for different values of $M$ , one
obtains the scattering phase shift (\ref{fieldeffrange}).
The phase shift of the underlying microscopic model
(\ref{potshift}) shows a lot of structure at the 
microscopic scale $k\sim a$. However, at low
energies $ka<<1$ it shows the same behavior (\ref{potshift0}) as the
baby field theory. 

Comparing (\ref{potshift0}) to (\ref{fieldeffrange}) relates
the effective field theoretic coupling $g$ to the microscopic
parameters:
\begin{equation}
g^2=\xi^2{16(2\pi)^3\Omega_d\nu(1-\nu^2)\over
\nu-\eta^{2(1-\nu)}+(1-\nu)\eta^2}.
\end{equation}
Notice that weak field theoretic coupling $g\to0$ corresponds 
at fixed $a,b$ to
the height of the barrier going to infinity ${\hat \lambda}\to\infty$
while the coefficient of the attractive component of the
potential goes to a finite limit ${\hat\gamma}\to
\pi\nu/(1-\eta^{2\nu})$. The opposite limit $g\to\infty$ corresponds
to vanishing ${\hat\gamma}$ and ${\hat\lambda}$ approaching
a finite negative constant. Thus in this latter limit the barrier
disappears.
\setcounter{equation}{0}
\section{Discussion}
Our crude two bit model illustrates the mechanism we have in
mind for dealing with the variable $P^+$ carried by lines in
light-front Feynman diagrams. Instead of explicitly summing
over each $P^+$, it is hoped that the tight-binding part
of the interaction potential will cause a collection of $M$
``bits'' with Newtonian mass $m$ to behave as a single
particle with Newtonian mass $Mm$. Of course, for this to
really work, the many body bound states must exhibit many
consistency conditions embodied in the fact that they
must act as a component of a relativistic field.

For $M$ larger than 2, it is not at all clear
for a generic field theory that a restriction
to only two body interactions will afford enough flexibility
to meet these conditions. For example, a 
one-dimensional many particle system with
the same attractive delta function interaction
between each pair is exactly soluble but has entirely
the wrong scaling behavior with large $M$. 

\begin{figure}
\centering
\centerline{
\psfig{figure=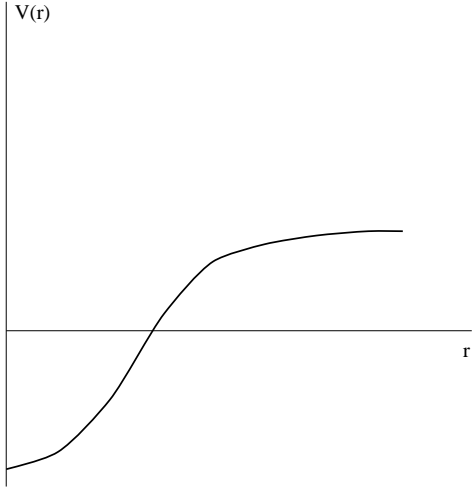,height=6.5cm}
}
\caption{Potential energy function for fundamental string.}
\label{stringpot}
\end{figure}

However, for large $N_c$ matrix field theories the prospects
are brighter. This is because, as shown in \cite{thornfock},
the dynamics of the the large $N_c$ limit can be mapped onto 
those of a linear chain on the light-front. The field quanta
or partons are in this limit ordered around a ring and only
nearest neighbors on the ring interact. In string-bit models
of fundamental string, all partons are ``wee'' and nearest neighbors
interact via a potential of the shape shown in Fig. \ref{stringpot}.
As is well-known \cite{thornmosc}, this sort of dynamics leads to precisely
the Nambu-Goto string. For a confining field theory like
QCD, however, the chain dynamics includes processes in which
the gluon quanta fuse and fission so that the number of
gluons is not conserved.
\input psfig
Let us recall how large $N_c$ gluon dynamics was formulated in
\cite{thornfock}. After discretizing $P^+$ in the usual way,
we can consider the dynamics of a glueball carrying $M$ units
of $P^+$. Then a state of the glueball can be described by
an $M$ component wave function, the $k$th component of
which describes a system of $k$ gluons and therefore depends on the
transverse coordinate, polarization, and number of $P^+$ units
of each gluon and is cyclically symmetric:
\begin{equation}
\Psi_k({\bf x}_1,i_1,M_1;\cdots;{\bf x}_k,i_k,M_k)=
\Psi_k({\bf x}_k,i_k,M_k;{\bf x}_1,i_1,M_1;\cdots;
{\bf x}_{k-1},i_{k-1},M_{k-1}),
\end{equation}
with $\sum M_k=M$. Gluon dynamics is then formulated as a set
of $M$ coupled Schr\"odinger equations of the schematic form
\begin{equation}
\left(\sum_{l=1}^k{{\bf p}_l^2\over 2mM_l}+V_{k,k}-E\right)\Psi_k=
-V_{k,k-2}\Psi_{k-2}-V_{k,k-1}\Psi_{k-1}-V_{k,k+1}\Psi_{k+1}
-V_{k,k+2}\Psi_{k+2}.
\end{equation}
The term $V_{k,k}$ is a sum of nearest neighbor interaction
potentials amongst the $k$ gluons described by $\Psi_k$. It
is actually a matrix differential operator because gluon  spin and $P^+$ can
be exchanged between the two neighbors. The coupling terms
on the r.h.s. take into account the possibility of a change
in the number of gluons. In each case these number changes 
respect the cyclic ordering. For example, by virtue of the
cubic Yang-Mills vertex, a pair of nearest neighbors can convert into a 
single gluon, and that gluon occupies the same spot on the chain
as the original pair. Similarly if a single gluon on the chain
converts to a pair, that pair's chain location is the same as
that of the original gluon. Because of this nearest neighbor
pattern, the processes just described are not unlike those
in the baby field theory described in Section 3. Just as we eliminated
the $M=2$ component in that case, we could imagine eliminating
all of the $\Psi_k$ for $k<M$, ending up with a horrific single
equation for the ``wee parton'' component $\Psi_M$. Such a
procedure looks hopelessly intractable.

\begin{figure}
\centering
\centerline{
\psfig{figure=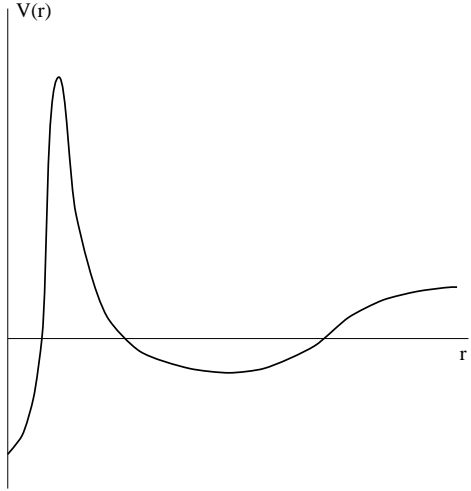,height=6.5cm}
}
\caption{Potential energy function for confining field theory.}
\label{qcdpot}
\end{figure}

Instead, we are suggesting in this article that by modifying the
terms in $V_{M,M}$ to have a potential shape indicated in
Fig. \ref{qcdpot}, one might do away with all the components
$\Psi_k$, $k<M$  accounting for their effects
as tunneling processes described by the 
new Schr\"odinger equation for $\Psi_M$.
The long distance attractive potential well
accounts for the stringy (confinement) behavior of a gluon chain and the
short distance attraction and barrier enable long-lived
tightly bound clusters of wee gluons which, we hope, act like 
valence gluons. To explore further this possibility, it
is probably not a good idea to try to derive this
new potential from gluon dynamics. 

This is because the new potential describes
an underlying theory different from QCD: the scale
$a$ is the scale at which gluons show
compositeness, for instance it could
be the distance scale of fundamental string. 
Rather, one should directly explore the underlying 
theory and try to test whether
it can reproduce QCD physics. Among these tests 
would be to see whether the right $M$ dependence can
come out of the gluon number changing transitions arising
from tunneling through the short range barrier. The nearest
neighbor interaction pattern of the large $N_c$ limit provides
a natural similarity between the conversion of a cluster
of varying size into two smaller clusters: the tunneling
process only involves the single link between
the two clusters regardless of the cluster size. Of course
larger clusters will have more inertia so the transition
amplitudes {\it will} depend on cluster size. Another
favorable circumstance is that the nearest neighbor pattern
will naturally make the clusters polymeric and therefore
stringy scaling laws are more likely. 

Although we have not taken into account the many body aspects
of this scheme for dealing with large $N_c$ QCD, we can at
least roughly understand why the limit of large `t Hooft coupling
entails a wee parton approximation. Referring back to our
baby field theory, it is the coupling $g^2$ that plays the
role of $N_cg^2$. We have seen that, in the large $g$ limit,
the barrier of Fig. \ref{qcdpot} disappears. Thus the
nearest neighbor interaction reverts to a simple potential well as
in Fig. \ref{stringpot} which wipes out the clustering effects
responsible for valence partons.

Finally, to bring this discussion full circle, we would like to
note that there is similar physics lurking in the AdS--QCD
connection, or more precisely in Polyakov's ``confining
string'' proposal \cite{polyakovconfine}. He suggests
that the coefficient $a(\phi)$ of $(\partial x)^2$ in the usual
world sheet action should depend on the Liouville field $\phi$. 
$a(\phi)$ then has the interpretation of a dynamical tension.
In \cite{gubserkp} $\phi$ is just the ``fifth'' dimension of
AdS$_5$. When such a world sheet dynamics is referred to
the light front, one finds the Hamiltonian
\begin{equation}
P^-=\int_0^{P^+}d\sigma{1\over2}[{\bf\cal P}^2
+a^2(\phi){\bf x}^{\prime2}+a(\phi)(\Pi_\phi^2+\phi^{\prime2})].
\end{equation}
With $\sigma$ discretized, the significance of $a^2(\phi)$
becomes a dynamical spring constant, and the quantum
dynamics of $\phi$ can be interpreted as a certain average over
variable spring constants. For harmonic oscillators,
averaging over variable spring constants is equivalent (dual) to
averaging over masses. But averaging over masses is what
is accomplished by our clustering of wee partons into
valence partons. What is not at all clear, of course, is whether 
the weightings of these averages have anything to do with one another.

\medskip
\noindent \underline{Acknowledgements}. I should like to thank Pierre Sikivie 
and Stan Brodsky for their helpful comments on the manuscript.

\bibliography{../larefs}
\bibliographystyle{unsrt}

\end{document}